\documentstyle[twocolumn,aps,prl,epsfig]{revtex}
\begin{document}
\draft
\wideabs{
\title{Method for Precision Test of Fine Structure Constant Variation with Optical Frequency References}
\author{J.R. Torgerson}
\address{Los Alamos National Laboratory, Physics Division, P-23, MS-H803, Los Alamos, NM 87545}
\date{\today}
\maketitle

\begin{abstract}

A new method for examining the possible space-time variation of the fine structure constant ($\alpha$) is
proposed. The technique uses a relatively simple measurement with an optical resonator to compare
atom-stabilized optical frequency references. This method does not require that the exact
frequency of each reference be measured, and has the potential to yield more than a 1000-fold improvement
in experimental sensitivity to changes in $\alpha$. A specific realization of an experiment using this
method is discussed which can approach the precision of ${\dot\alpha/\alpha}\!\sim\!10^{-18}/\tau$, where
$\tau$ is the measurement time. Moreover, for this specific realization, a measurement of
${\dot\alpha/\alpha}\!\sim\!10^{-15}/{\rm yr}$ as a near-term goal is realistic.

\end{abstract}
}

As space-time variation of fundamental constants is implicitly precluded from widely accepted physical
theories, a search for such variation is a probe for physics beyond our current understanding. Although a
measured variation would not necessarily disprove long-standing theories such as general relativity,
quantum electrodynamics and others, it would be evidence of deeper structure in nature. Moreover, the
sensitivity of these searches can best the most accurate physical measurements. 

The notion that our physical description of nature is incomplete is supported with empirical
evidence. Two relevant examples are: 1) the set of constants termed `fundamental' are more numerous
than, and hence overdefine the absolute scales of length, mass, electric charge, {\it etc}., and 2) we
have no verified theory that unifies gravity with the other basic forces. 

Numerous theories currently under investigation have been devised to address these issues which either
allow or necessitate space-time variation of fundamental constants such as the fine structure constant
($\alpha$)~\cite{DiracLNH,Teller48,Barrow87,Damour94,Freund82,Marciano84,Carroll98}. 

A number of workers have experimentally determined limits for the time dependence of the fine
structure
constant~\cite{Schlyakter76,Sisterna90,Damour96,Webb99,Ivanchik99,Webb01,Turneaure76,Prestage95,Sortais00}. 
The sources of information used by these workers can be divided into two categories: observational data
and laboratory measurements. Currently the most constraining limits for $\dot\alpha$~($\equiv
d\alpha/dt$) come from observational data. The first result reported in Table~\ref{AlphaDotLimits} was
obtained from analysis of the natural fission reaction that took place in Oklo, Gabon about
$\!2\!\cdot\!10^9$ years ago. The next result was obtained through analysis of astrophysical data and
reports the first observation of a $4\,\sigma$ deviation from $\dot\alpha=0$. While this result will
undoubtedly recieve much scrutiny, it has highlighted the importance of other measurements to achieve the
same accuracy.

The observational methods have the potential advantage of time averaging over billions of years when
searching for a linear drift in $\alpha$. This advantage combined with the ability to look for both
ancient and large-scale spatial variation in $\alpha$ make the observational methods valuable for the
future of this research. However, these methods are not without difficulties. Most importantly, the
interpretation of these data is model dependent~\cite{Sisterna90} and the data may reflect alterations
by unknown mechanisms~\cite{Webb99}.

The final two values in Table~\ref{AlphaDotLimits} are the results of laboratory measurements.
Laboratory searches for $\dot\alpha$ have several benefits over observational methods: The measurements
can be made in a carefully controlled environment, are not model dependent and can be reproduced
independently. Laboratory measurements have an additional benefit of being sensitive to both the
present-day value of $\alpha$ and oscillations that may leave the integrated value of $\alpha$ unchanged
over cosmic time scales.

The two laboratory limits reported in Table~\ref{AlphaDotLimits} are the results of microwave
time standards comparisons. The basis for these comparisons in a search for nonzero
$\dot\alpha$ is that configuration energies of different atoms depend differently on
$\alpha$. For the microwave standards referenced in Table~\ref{AlphaDotLimits}, the change in
frequency of the ``clock'' transition for nonzero $\dot\alpha$ scales to lowest order in  $\alpha$
as $(\alpha Z)^x$, where $Z$ is the atomic number and $x\sim2$.

In general, comparison of two atom-stabilized time standards can yield a limit on the fractional change
in $\alpha$ during the measurement time that is on the order of, but not better than, the fractional
stability of the least precise of the standards being
compared~\cite{Sisterna90,Prestage95,Turneaure76,Sortais00,Dzuba99}. Table~\ref{ClockLimits} illustrates
the current status of various time standards in existence today. For these standards, the current limit for
$\dot\alpha/\alpha$ could see an improvement of an order of magnitude in the forseeable future. 
However, it is likely that further improvements on this limit will be constrained by magnetic field
shifts or collisional shifts.

It has long been understood that certain narrow atomic transitions whose frequencies are in the optical
band could be used to create frequency references, and possibly time standards, with far improved
stability over the current state-of-the-art~\cite{Dehmelt73}. The most significant obstacle in performing
an experiment to compare frequency references in which one of the references is optical, or to use an
optical frequency reference as a time standard, is the optical frequency metrology. Yet with significant
effort, some workers have discovered clever methods to relate optical frequencies to directly measurable
microwave standards~\cite{OpticalComb99}, such as one of the microwave standards shown in
Table~\ref{ClockLimits}. Not only are these measurements difficult, but the accuracy of the
optical frequency metrology depends on the stability of the microwave standard. If the intrinsic
stability of the optical reference is better than the microwave standard, this comparison adds noise
into the measurement and reduces the available information about the optical frequency reference. Using
this method to compare two or more such references would yield far less information than could be
gained in principle.

It is therefore significant that the method proposed here makes comparisons only between optical
frequency references without the potentially noisy intermediate step of comparing each to a microwave
standard. The necessary comparison can be made directly between the references with a
simultaneously-resonant optical cavity, or optical comparison resonator (OCR).

\begin{figure}
  \begin{center}
    \epsfxsize=.8\columnwidth
    \epsfbox{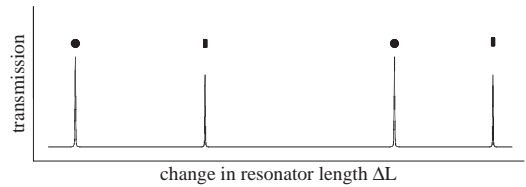}
  \end{center}
  \footnotesize
  \caption{Resonator transmission {\it vs.} change in resonator length $\Delta
L$: Different symbols mark transmission peaks for incident fields of different frequencies $\nu_i$.
Separation between peaks with similar symbols is $\nu_{fsr}$ ($\Delta\nu_i\!=\!-(\nu_i/L)\Delta L$).}
 \label{ComparisonResonator}
\end{figure}

To illustrate, consider narrow-band sources of coherent optical radiation with frequencies $\nu_1$
and $\nu_2$ that are referenced to metastable transitions of suitable atoms or ions. A plot of the
transmitted intensity of an optical resonator illuminated with these optical fields as a function of
resonator length would resemble Fig.~\ref{ComparisonResonator} where the maximum separation (in Hz)
between the resonance peaks of the two fields is one-half of the free-spectral-range
($\nu_{fsr}=c/L$, where $c$ is the speed of light and $L$ is the round-trip length of the resonator).
Obviously, if the length of the resonator is stabilized by maintaining a resonance with one of the optical
fields, the other field can be shifted into resonance simultaneously with this OCR with
an acousto-optic modulator (AOM) or similar device where the frequency of the RF oscillator driving
the AOM is less than
$\nu_{fsr}/2$. Information about changes in the relative frequencies of the atom-stabilized references
can be gathered by monitoring the frequency of the RF signal.

This is important because {\it it is sufficient to measure changes in the difference frequencies of
relatively stable atomic references to measure changes in $\alpha$}. Moreover, an experiment to measure
changes in $\alpha$ with atom-stabilized frequency references does not require that the frequencies of
the references be known exactly, or even that their difference frequencies be known exactly.

Calculations were made recently, with a combination of the relativistic Hartree-Fock model and many-body
perturbation theory, of the relativistic energy level corrections as an expansion in powers of $\alpha$
for In$^+\!$ and Tl$^+\!$~\cite{Dzuba00}. The results can be expressed in the form 
$\dot\omega\!=\!\beta(\dot\alpha/\alpha)$. For the
$^1\!S_0\!\leftrightarrow\!{}^3\!P_0$ optical transition in In$^+\!$,
$\beta\!=\!2.6\!\cdot\!10^{14}$\,Hz. While for the same transition in Tl$^+\!$,
$\beta\!=\!12\!\cdot\!10^{14}$\,Hz. The large difference between In$^+\!$ and Tl$^+\!$ is due to the
increased magnitude of the relativistic corrections for larger atomic number. The scaling rule for this
effect is not simple but the magnitude of the corrections is expected to increase as $(\alpha{}Z)^x$.
With $Z\!=\!49$ and 81 for In and Tl respectively, the value 3.0 for $x$ agrees quite well
with the $\beta$ values calculated from~\cite{Dzuba00}.

Assuming the In$^+\!$ and Tl$^+\!$ references possess the same stability, the accuracy to which 
$\dot\alpha/\alpha$ can be determined is $3.0\,\sigma$ where $\sigma$ is the fractional uncertainty of
the In$^+\!$ and Tl$^+\!$ frequency references. As noted below, $\sigma$ decreases as $t^{-1/2}$.
Because, to first order, a nonzero $\dot\alpha$ would change linearly with time, the sensitivity of the
proposed experiment to $\dot\alpha$ improves as $\approx\!t^{-3/2}$. At such time that the precision of
the frequency references becomes limited by white noise, the sensitivity improves as $\approx\!t^{-1}$.

\begin{figure}
  \begin{center}
    \epsfxsize=\columnwidth
    \epsfbox{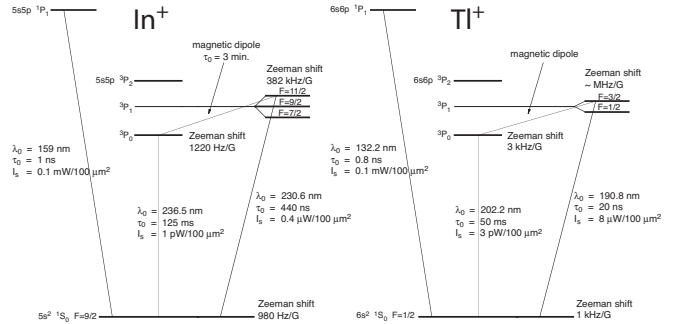}
  \end{center}
  \footnotesize
  \caption{Grotrian diagrams for In$^+\!$ and Tl$^+\!$.}
  \label{GrotrianDiagrams}
\end{figure}

A trapped ion time or frequency reference can offer significant improvements in stability over neutral
atom references. Evidence of this appears in Table~\ref{ClockLimits} which shows that the Hg$^+\!$
reference stability is limited by magnetic field fluctuations and not by uncontrollable collisional
effects. Table~\ref{IonTrapVirtues} presents several virtues of a trapped ion standard. For In$^+\!$ in
a harmonic trap with an easily obtainable frequency
$\omega_i/2\pi\!=\!1$\,MHz~\cite{PaulTrap,Yu95}, the extent of the motional state of the ion can be
described by $\langle{}z^2\rangle^{1/2}\!\approx\!7$\,nm~\cite{Wineland79}. The transition of interest 
$^1\!S_0\!\leftrightarrow\!{}^3\!P_0$ (Fig.~\ref{GrotrianDiagrams}) has  $\lambda/2\pi\approx40$\,nm so
that the first-order Doppler shift is clearly eliminated. Similarly, the second-order Doppler shift for
In$^+\!$ can be  $\Delta\omega_D/\omega_0<10^{-19}$. Collisional shifts can be reduced to
$\Delta\omega_c/\omega_0<10^{-19}$ with readily available vacuum technology.

Field shifts for In$^+\!$ and Tl$^+\!$ can be reduced below $10^{-18}$ fractional frequency using
standard techniques~\cite{Dehmelt73,Walther94}. Table~\ref{IndiumVirtues} shows the field shifts of
In$^+\!$; values for Tl$^+\!$ are similar. Magnetic fields can be reduced in a small
experimental apparatus to the level of micro-Gauss with passive shielding, and the Stark shifts of the
cooled indium ion in the previously described trap are
$\sim\!10^{-20}\!\cdot\omega_0$.

Moreover, optical atomic references offer a significant advantage over microwave references in short-time
stability. For an atomic reference limited by quantum statistics and using the Ramsey's method of
separated oscillatory fields~\cite{RamseyClock}, the two-sample Allan variance takes the form
$\sigma=1/\omega_0\sqrt{N T_R t}$~\cite{QuantumSigma}, where $\omega_0$ is the center frequency of the
reference, $N$ is the number of (uncorrelated) atoms contributing to the signal, and $T_R$ is the
temporal extent of a single measurement, and $t$ is the total integration time.  Optical
references offer an improvement over microwave references by $\sim\!10^5$ in $\omega_0$.

There are several issues to be considered regarding the OCR for the sensitive measurement proposed in
this manuscript. If a resolution of better than 1\,Hz can be obtained, the OCR will not limit the
accuracy to which $\dot\alpha$ can be determined with this method. This criterion suggests that the
resonator linewidth should be under 10\,kHz~\cite{SubHertz}. Also, to guarantee a simultaneous resonance
with the use of an AOM, the free-spectral-range of the OCR should be less than twice the AOM tuning
range of a few hundred MHz. These criteria can be met for a multiply-resonant cavity with
currently available optics technology.

There are shifts and broadening mechanisms other than those discussed above that can lead to systematic
errors in this type of measurement. For example, Doppler shifts between the ions and the OCR could cause
an anomalously positive result. These shifts arise largely from acoustical noise which exists mainly
below 1\,kHz. A first-order Doppler shift caused by a 30\,Hz vibration with an amplitude of 20\,nm will
shift the OCR line center by about 1\,Hz. This is small compared with the expected $\sim\!10^3-10^4$\,Hz
OCR spectral width and the long-time effect of this shift is similar to a small broadening of the
OCR linewidth. It does, however, obscure any information about frequency difference changes for
times on the order of the inverse vibrational frequency. At a similar noise level, the second-order
Doppler shift is negligible: More specifically, the amplitude of 30\,Hz vibrations would have to be
nearly a meter to effect a similar shift. 

A change in the pressure will change the index of refraction of the air between the OCR
mirrors and will alter the resonance condition. The low pressure limit of the index of refraction
of standard air~\cite{AirIndex} can be used to estimate this effect. The result is
$\Delta n\!=\!4\!\cdot\!10^{-7}\Delta p/torr$. Pressure changes of $10^{-8}$\,torr will shift the
resonance by about 1\,Hz. This effect can be reduced to $<10^{-3}$\,Hz by placing the resonator in a
vacuum of $10^{-11}$\,torr. At this pressure, $\Delta n\!=\!-1\!\cdot\!10^{-19}\Delta T/K$ where $\Delta
T$ is the temperature change.

Unless the intracavity power of the OCR is stabilized, the mirror coatings may
heat and cool causing additional broading of the resonance~\cite{SubHertz}. This effect can be
mitigated by using low power and stabilizing the OCR throughput.

A cesium reference or similar atom-stabilized oscillator will be needed to synthesize the RF drive for
the frequency shifters used to achieve the simultaneous resonances in the OCR.
With a fractional stability of better than $10^{-12}$ and a synthesized frequency of $10^8$\,Hz, the
uncertainty caused by this type of oscillator in this measurement is less than $10^{-19}$. Also, the
frequency of any atom-stabilized reference will change for a nonzero $\dot\alpha$. This effect
will not reduce the sensitivity of the proposed measurement as any such shift will be reduced by the
factor of $10^5$ frequency ratio of the optical and microwave atom-stabilized oscillators.

A linear Paul trap is a convenient device to trap the indium and thallium ions simultaneously.  There
are several advantages to this approach: The first advantage is that the indium cooling and
interrogation transition is narrow enough to allow cooling directly to the motional ground state. As the
ions' motion is coupled through their mutual repulsion, the indium ion can be used to sympathetically
cool an In$^+\!$-Tl$^+\!$ two-ion crystal to the motional ground state. The similar transition in
thallium is too broad to cool to this final motional state independently. Secondly, the ion separation
will be about $20\,\mu{\rm m}$ ensuring that both ions are exposed to the same low frequency electric and
magnetic fields.

A potential limitation of this technique is that neither ion will be at the zero-field point along the
trap's axis so that both ions will be exposed to a small RF electric field that can cause a quadratic
Stark shift of the metastable transition. The most likely cause for this unfavorable condition is small
misalignments of the trap electrodes. The magnitude of the electric field to which they could be exposed
can be approximated by $E_d\approx(V_0/r_0)\!\cdot\!(d/r_0)$ where $d$ is the distance of the ions from the
ideal trap center. For typical parameters, $E_d\!\ll\!1\,{\rm V/cm}$ and the fractional shift magnitude
for each ion is $\ll\!10^{-18}$ (with the same sign). Moreover, micromotion of this magnitude is easily
detectable.

Due to the similarity of the $^1\!S_0\!\leftrightarrow\!{}^3\!P_0$ transition frequencies of In$^+\!$ and
Tl$^+\!$, an order of magnitude of insensitivity to Doppler broadening, pressure shifts and mirror
heating effects is gained over the limits discussed above. Additional benefits for comparing two IIIA
ions are gained due to the similarity of the electronic configurations. For example, frequency difference
measurements between In$^+\!$ and Tl$^+\!$ are less sensitive to some field shifts.

In conclusion, a test of possible space-time variation of the fine structure constant based on
comparisons of optical frequency references is proposed. The required measurement of a change in the
{\it relative} frequencies of the atom-stabilized references can be performed by employing a
multiply-resonant optical cavity. Moreover, exact measurements of the reference frequencies are not
necessary for this experiment which greatly reduces the complexity of this experiment from that of
atomic clock comparisons proposed for the same purpose. The choice of In$^+\!$ and Tl$^+\!$ ions
simultaneously confined in a linear trap allows the In$^+\!$-Tl$^+\!$ crystal to be cooled to the
motional ground state with a single laser. This configuration does not limit the ultimate relative
stability of corresponding frequency references. By using two group IIIA ions, a limit of
${\dot\alpha/\alpha}\!\sim\!10^{-18}/\tau$ can be approached, where $\tau$ is the time over which the
frequency references are compared. A measurement of
${\dot\alpha/\alpha}\!\sim\!10^{-15}/{\rm yr}$ as a near-term goal is realistic and would provide the
finest laboratory resolution for $\dot\alpha$.

Note that a future refinement to this experiment could include other simultaneously
trapped ions and comparisons of the resulting optical frequency references. The different dependencies on
$\alpha$ would allow weaker requirements for measuring and eliminating systematic effects and would be
especially important for quantifying any nonzero $\dot\alpha$, should one exist.

For their comments and discussions, I would like to acknowledge W.T. Buttler, R.J. Hughes, S.K.
Lamoreaux, W.N. Nagourney and V.D. Sandberg.

\begin{table}
  \begin{tabular}{rlr}
    \multicolumn{1}{c}{\it limit for $\dot\alpha/\alpha$}&\multicolumn{1}{c}{\it
    method}&\multicolumn{1}{c}{\it ref.}\\\hline
    $10^{-15}/{\rm yr}$&Oklo reaction&\cite{Schlyakter76,Sisterna90,Damour96}\\
    $(-1.0\pm0.2)\!\cdot\!10^{-15}/{\rm yr}$&quasar spectra&\cite{Webb01}\\\hline

    $3.7\!\cdot\!10^{-14}/{\rm yr}$&H--maser\,$\rightleftharpoons$\,Hg$^+$clock&\cite{Prestage95}\\
    $(1.8\pm2.7)\!\cdot\!10^{-14}/{\rm yr}$ &Rb fountain\,$\rightleftharpoons$\,Cs fountain&
      \cite{Sortais00}\\\hline  
  \end{tabular}
  \footnotesize
  \caption{Current limits placed upon $\dot\alpha/\alpha$.}
  \label{AlphaDotLimits}
\end{table}

\begin{table}
  \begin{tabular}{lrlr}
    \multicolumn{1}{c}{\it type}&\multicolumn{1}{c}{\it Allen variance}&
    \multicolumn{1}{c}{\it limiting factor}&\multicolumn{1}{c}{\it ref.}\\\hline
    Cs fountain&$6\!\cdot\!10^{-16}$&quantum statistics&\cite{Santarelli99}\\
    Hg$^+$&$3.4\!\cdot\!10^{-15}$&applied field Zeeman shifts&\cite{WinelandHg98}\\
    H--maser&$2\!\cdot\!10^{-15}$&collisional perturbations&\\
  \end{tabular}
  \footnotesize
  \caption{Fractional uncertainty in the oscillator frequency of current time standards.}
  \label{ClockLimits}
\end{table}

\begin{table}
  \begin{tabular}{lll}
    \multicolumn{3}{l}{First-order Doppler shift for $\langle z^2\rangle\ll{1/k^2}$:}\\
    \multicolumn{1}{l}{\hbox spread0.25\columnwidth{}}&\multicolumn{1}{l}{}&
    \multicolumn{1}{l}{${\Delta\omega_d\over\omega_0}=0$}\\
    \multicolumn{3}{l}{Second-order Doppler shift:}\\
    \multicolumn{1}{l}{}&\multicolumn{1}{l}{${\Delta\omega_D\over\omega_0}={3kT\over2mc^2}$}&
    \multicolumn{1}{l}{${3k\over2c^2}=1\!\cdot\!10^{-19}\;{{\rm amu}\over\mu{\rm K}}$}\\
    \multicolumn{3}{l}{Collisional broadening (optical transition):}\\
    \multicolumn{1}{l}{}&\multicolumn{1}{l}{}&\multicolumn{1}{l}{${\Delta\omega_c\over\omega_0}%
    \sim{10^{-8}\!\cdot\!P\;[{\rm torr}]}$}\\
    \multicolumn{3}{l}{No transit-time broadening}\\
    \multicolumn{3}{l}{No loss of signal over time}\\
  \end{tabular}
  \footnotesize
  \caption{Facts of interest for ion-trap standards.}
  \label{IonTrapVirtues}
\end{table}

\begin{table}
  \begin{tabular}{lll}
    \multicolumn{3}{l}{Zeeman shift:}\\
    \multicolumn{1}{l}{\hbox spread0.1\columnwidth{}}&
    \multicolumn{1}{l}{${\Delta\omega_h\over2\pi{}H}=0.24\,{{\rm mHz}\over\mu{\rm G}}$}&
    \multicolumn{1}{l}{${\Delta\omega_h\over\omega_0}=3\!\cdot\!10^{-20}\!\cdot\!H\;
      [\mu{\rm G}]$}\\
    \multicolumn{3}{l}{Quadratic Stark shift:}\\
    \multicolumn{1}{l}{}&
    \multicolumn{1}{l}{${\Delta\omega_s\over2\pi{}E^2}\sim{\rm mHz\over\left(V/cm\right)^2}$}&
    \multicolumn{1}{l}{${\Delta\omega_s\over\omega_0}\sim10^{-21}\!\cdot\!T\;[\mu{\rm K}]$}\\
    \multicolumn{3}{l}{Electric quadrupole shift:}\\
    \multicolumn{1}{l}{}&\multicolumn{1}{l}{$\sim\!0$ for $^1\!S_0\!\leftrightarrow\!{}^3\!P_0$}&
    \multicolumn{1}{l}{($10^6$ less than $^1\!S_0\!\leftrightarrow\!{}^3\!P_1$)}\\
  \end{tabular}
  \footnotesize
  \caption{Field shifts of interest for an In$^+$ reference.}
  \label{IndiumVirtues}
\end{table}

\end{document}